\documentclass[aps,pra,twocolumn,amsfonts,amssymb,amsmath,floatfix, superscriptaddress]{revtex4}

\usepackage{graphicx}
\usepackage{color}
\usepackage{subfigure}
\usepackage{bm}
\usepackage{epstopdf}
\usepackage{comment}
\usepackage{float}
\usepackage{bbold}
\usepackage{braket}

\begin{document}
\title{Strong correlation effects in a two-dimensional Bose gas with quartic dispersion}

\author{Juraj Radi\'{c}}
\affiliation{Joint Quantum Institute and Condensed Matter Theory Center, Department of Physics, University of Maryland, College Park, Maryland 20742-4111, USA}

\author{Stefan S. Natu}
\affiliation{Joint Quantum Institute and Condensed Matter Theory Center, Department of Physics, University of Maryland, College Park, Maryland 20742-4111, USA}

\author{Victor Galitski}
\affiliation{Joint Quantum Institute and Condensed Matter Theory Center, Department of Physics, University of Maryland, College Park, Maryland 20742-4111, USA}
\affiliation{School of Physics, Monash University, Melbourne, Victoria 3800, Australia}


\begin{abstract} 
Motivated by the fundamental question of the fate of interacting bosons in flat bands, we consider a two-dimensional Bose gas at zero temperature with an underlying quartic single-particle dispersion in one spatial direction. This type of band structure can be realized using the NIST scheme of spin-orbit coupling [Y.-J. Lin, K. Jim\'{e}nez-Garcia, and I. B. Spielman, Nature \textbf{471}, 83 (2011)], in the regime where the lower band dispersion has the form $\varepsilon_{\bm{k}} \sim k_{x}^{4}/4+k_{y}^{2}+\ldots$, or using the shaken lattice scheme of Parker \textit{et al.} [C. V. Parker, L.-C. Ha and C. Chin, Nature Physics \textbf{9}, 769 (2013)]. We numerically compare the ground state energies of the mean-field Bose-Einstein condensate (BEC) and various  trial wave-functions, where bosons avoid each other at short distances.  We discover that, at low densities, several types of strongly correlated states have an energy per particle ($\epsilon$), which scales with density ($n$) as $\epsilon \sim n^{4/3}$, in contrast to $\epsilon \sim n$ for the weakly interacting Bose gas. These competing states include a Wigner crystal, quasi-condensates described in terms of properly symmetrized fermionic states, and variational wave-functions of Jastrow type. We find that one of the latter has the lowest energy among the states we consider. This Jastrow-type state has a strongly reduced, but finite condensate fraction, and true off-diagonal long range order, which suggests that the ground state of interacting bosons with quartic dispersion is a strongly-correlated condensate reminiscent of superfluid Helium-4. Our results show that even for weakly-interacting bosons in higher dimensions, one can explore the crossover from a weakly-coupled BEC to a strongly-correlated condensate by simply tuning the single particle dispersion or density.

\end{abstract}

\maketitle

\section{Introduction}

One of the most remarkable advances in ultra-cold atomic gases in the recent years has been the ability to engineer at will, dispersions with single-particle degeneracies or almost completely flat bands. For example, optical superlattices have been used to generate honeycomb, Kagom\'e and Lieb lattice geometries~\cite{Honeycomb2012,Kagome2012}. Lattice shaking~\cite{Esslinger2014,Chin2013} and Raman assisted tunneling in real and spin space has been used to realize spin-orbit coupling (SOC)~\cite{Spielman2011,OurNature}, synthetic vector potentials, and subsequently topological bands~\cite{Bloch2013,Ketterle2013}. Attention has now turned to studying the interplay between these non-trivial single-particle band structures, spin and interactions, which paves the way to accessing a rich variety of phases such as skyrmion lattices~\cite{Pfleiderer2006}, integer and fractional Chern insulators~\cite{Haldane1988,Bernevig2011}, Wigner crystals \cite{Berg2012}, and other exotic states~\cite{WenBook}. Here we present a variational study of a low-density, $2$D Bose gas at zero temperature in a dispersion which is quartic in one direction, and which can be realized experimentally~\cite{Spielman2011, Chin2013}.

An interesting  example of non-trivial interplay between single-particle degeneracies and interactions is a $2$D  Rashba SOC gas. Here the low-energy dispersion has an infinite ring degeneracy in momentum space, and the
the density of states has the form $dn/dE \sim E^{-1/2}$, typical for $1$D systems. At low densities, atoms sample the ring degeneracy and interesting physics emerges. The consequences of this were first explored by Berg, Rudner and Kivelson \cite{Berg2012} in the context of Fermi gases. They observed that while the kinetic energy delocalizes the particles over the Rashba ring, atoms can minimize the short range interaction energy by localizing in momentum space. This competition produces a plethora of possible symmetry broken ground-state phases ranging from Wigner crystals to ferromagnetic nematic states.

Even more interesting and perhaps less understood is the fate of \textit{bosons} in single-particle degeneracies. On the one hand, by developing fermionic correlations, bosons can completely avoid (spinless case) or suppress (spinful case) short-range repulsive interactions, but such a state is spread out in momentum space. The kinetic energy cost associated with this spreading is parametrically lower in flat bands, and one can expect a regime of densities where \textit{fermionized} wave functions have lower energy than a mean-field condensate. The key theoretical challenge in addressing this question is that single particle degeneracies enhance fluctuation effects, rendering mean-field theory invalid, and Quantum Monte Carlo usually suffers from a sign problem, and can only study small system sizes. Progress has to be made either by guessing trial wave-functions or using field theoretical methods which capture the low energy dynamics. For Rashba SOC~\cite{SOBEC} and moat bands, Sedrakyan \textit{et al.}~\cite{Sedrakyan2012,Sedrakyan2014} have proposed a composite-fermion description, which spontaneously breaks time reversal and parity symmetry and has lower energy than the weak-coupling BEC. Spinless bosons in quartic bands of the form $\varepsilon_{\bm{k}} \sim \bm{k}^4$, were studied recently using field theoretic techniques by the authors of Refs.~\cite{QiZhou2014,Miao2015}, who proposed that condensation is strongly suppressed in favor of a liquid with algebraically decaying spatial correlations.


Motivated by experiments, we address the question of fermionization versus Bose condensation in a $2$D Bose gas in the NIST SOC  \cite{Spielman2011} or Chicago shaken lattice scheme \cite{Chin2013}, where the dispersion can be tuned to take the form $\varepsilon_{\bm{k}} \sim k_{x}^{4}/4+k_{y}^{2}+...$. We compare the energy of the mean-field Bose condensate to several trial many-body states, summarized in Table~\ref{tab1}: (i) a Wigner crystal 
(ii) the absolute value and the square of the Fermi-sea wave-function 
(iii) the absolute value and the absolute value squared 
of the $\nu = 1$ Laughlin state (proposed by the authors of ~Refs.~[\onlinecite{Sedrakyan2012,Sedrakyan2014}]) 
and (iv) the Jastrow ansatz \cite{Rossi2014}. While all the wave-functions (i)-(iv) have an energy per particle which scales as $\epsilon \sim n^{4/3}$ in the low-density limit (to be precisely defined below), and are thus energetically favorable over the mean-field condensate ($\epsilon \sim n$), we find that the trial wave-function with the lowest energy is of Jastrow type, and has finite condensate fraction and true long-range order. 

\section{The model}

We study a two-dimensional (pseudo)spin-$1/2$ Bose system with spin-orbit coupling, which was experimentally
realized at NIST~\cite{Spielman2011}:  
\begin{equation}
H_{\rm soc}(\bm{k})=\frac{\hbar^2 \bm{k}^2}{2m} + \frac{\hbar^2 k_L}{m} k_x \sigma_z + \frac{\Omega_R}{2} \sigma_x,
\label{H}
\end{equation}
where $k_L$ is the Raman laser wave-vector, $\Omega_R$ is the Raman coupling strength, and $\sigma_{x,y,z}$
are Pauli matrices. The spectrum of the Hamiltonian has two bands, for $\Omega_R < 4 E_{R}$, where 
$E_{R}=\hbar^2 k_{L}^{2}/2m$ is the laser recoil energy, the lower band has two degenerate minima, 
while for $\Omega_R \geq 4 E_{R}$ it has a single
minimum at $\bm{k}=0$ \cite{Spielman2011}. While the dispersion around each minimum is parabolic,
at $\Omega_R = 4 E_{R}$, the dispersion in the $x$-direction develops a quartic structure. In the case of a Bose gas,
this gives rise to interesting behavior at low densities, which is the main topic of the paper. From
now on, we will be interested only in the $\Omega_R = 4 E_{R}$ case. We remark that while the Rabi coupling term explicitly breaks physical time-reversal symmetry, this Hamiltonian has an additional $Z_{2}$ symmetry associated with the transformation $\ket{k_{x}, \uparrow} \rightarrow \ket{-k_{x}, \downarrow}$. 

Expressing energy in units of $E_{R}$ and momentum (length) in units of $\hbar k_L$ ($1/k_L$), the dimensionless single-particle Hamiltonian reads: 
%
\begin{equation}
H_k=\bm{k}^2 + 2 k_x \sigma_z + 2 \left( \sigma_x + \mathbb{1} \right).
\label{H_spinful}
\end{equation}
We choose the energy offset such that the minimum of the lower band is at zero energy. 

We assume the interactions between particles are described by a spin-independent contact potential (in units
of $E_R$ and $1/k_{L}$)
\begin{equation}
V_{\rm int} (\bm{r}_1-\bm{r}_2)= g \ \delta(\bm{r}_1-\bm{r}_2) \ \mathbb{1}_{\sigma_1 \otimes \sigma_2},
\label{V_int}
\end{equation}
where $g=2mU_{0}/\hbar^2$ ($U_{0} > 0$ is the contact interaction strength and in a quasi-$2$D regime 
$U_{0}=2\sqrt{2\pi} \hbar^2 a/(m a_z)$, where $a$ is a $3$D scattering length and $a_z$ is the confinement 
length in $z$ direction~\cite{Petrov2000}), $\mathbb{1}_{\sigma_1 \otimes \sigma_2}$ is a unit operator in the space of two spins,
$\mathbb{1}_{\sigma_1 \otimes \sigma_2}=\sum_{s_1,s_2} \lvert s_1 s_2 \rangle \langle s_1 s_2 \rvert$,
where $s_j \in (\uparrow, \downarrow)$.
In reality, the interactions are typically spin-dependent, however our results are insensitive
to spin dependence. We emphasize that throughout, we focus on the regime of weak interactions, but nonetheless find interesting ground states by engineering the single-particle dispersion. 

%
%
The spectrum of $H_k$ is $\varepsilon_{\pm}=\bm{k}^2 \pm 2\sqrt{k_{x}^{2}+1}+2$ and the lower-band
energy can be expanded around $\bm{k}=0$ as $\varepsilon_{-}(\bm{k}) = k_{x}^{4}/4 + k_{y}^{2}+...$.
The lower-band eigenstates of $H_k$ are
\begin{equation}
\begin{bmatrix}
s_{\uparrow}(\bm{k}) \\
s_{\downarrow}(\bm{k})
\end{bmatrix}
=\mathcal{N}_k 
\begin{bmatrix}
k_x-\sqrt{1+k_{x}^{2}} \\
1
\end{bmatrix},
\label{spin_eig}
\end{equation}
where $\mathcal{N}_k=\left[ 1+(k_x-\sqrt{1+k_{x}^{2}})^2 \right]^{-1/2}$ is the normalization factor.
Notice that at low densities ($n \ll k_{L}^2$, in original units), particles occupy only the states close to the minimum
of the band, i.e. the width of the momentum distribution $\Delta k_x \rightarrow 0$ as $n \rightarrow 0$.
In that case, Eq.~(\ref{spin_eig}) reduces to 
$\left[ s_{\uparrow}(\bm{k}) \ s_{\downarrow}(\bm{k}) \right] = [-1 \quad 1]/\sqrt{2}$, and spin eigenstates become
(approximately) momentum independent. The gas then becomes effectively spinless, and is 
described by the Hamiltonian:
\begin{equation}
\mathcal{H}_{\bm{k}}=\frac{1}{4} k_{x}^{4}+k_{y}^{2},
\label{H_spinless}
\end{equation}
with $\mathcal{V}_{\rm int} (\bm{r}_1-\bm{r}_2)= g \ \delta(\bm{r}_1-\bm{r}_2)$. Such a Hamiltonian can be directly realized using the shaking lattice scheme of Parker \textit{et al.} \cite{Chin2013}.

In the first part of this paper, we focus on the physics of the effective Hamiltonian Eq.~(\ref{H_spinless}) above, and then show that our conclusions remain unchanged even after the inclusion of spin (corresponding to the NIST scheme \cite{Spielman2011}). 


\section{Bogoliubov mean-field theory}
\label{sec3}

We start by considering the most conventional description of a $2$D Bose gas at zero temperature, namely the Bogoliubov mean-field description. 
The main assumption in Bogoliubov's approach is that the majority of particles are condensed 
in $\bm{k}=0$ state, and
others occupy $\bm{k} \neq 0$ states in the vicinity. Repulsive interactions deplete the condensate~\cite{Leggett2001}, and at the mean-field level, the 
energy per particle is given by $\epsilon = gn/2$, where $n$ is the density. The condensate depletion is readily found to be~\cite{PethickSmith}:

%
\begin{equation}
n_{\rm ex}=\frac{1}{V} \sum_{\bm{k} \neq 0} \frac{1}{2} \left( \frac{\varepsilon_{\bm{k}} + g n_0}
{\sqrt{\varepsilon_{\bm{k}}^2 + 2\varepsilon_{\bm{k}} g n_0}} - 1 \right),
\label{n_ex}
\end{equation}
where $\varepsilon_{\bm{k}}$ is a single-particle dispersion, $n_{\rm ex}$ is the density of depleted particles,
$n_0$ is the condensate density, and $g$ is the interaction strength.
The behavior of the integral in (\ref{n_ex}) is usually a good indication of the fate of a BEC:
for zero temperature, $2$D and $3$D systems with a parabolic dispersion, the integral is convergent, fluctuations do not destroy long-range order. In $1$D, it diverges, signaling the absence of true long-range order.

In our case, $\varepsilon_{\bm{k}} = k_{x}^{4}/4 + k_{y}^{2}$, the integral is convergent:
$n_{\rm ex}=3.854 (g n_{0})^{3/4}$ (dimensionless variables). However, the ratio of the number
of excited and condensed particles $n_{\rm ex}/n_0 \sim n_{0}^{-1/4}$ (in a usual $2$D parabolic
case $n_{\rm ex} \sim n_0$) shows that the Bogoliubov
approach breaks down at low densities. 
This suggests that in the low-density limit, the ground state 
is \textit{qualitatively} different from a mean-field condensate.

In the $3$D case, $\varepsilon_{\bm{k}} = k_{x}^{4}/4 + k_{y}^{2} + k_{z}^{2}$, the ratio of the number of 
excited and condensed particles is $n_{\rm ex}/n_0 \sim n_{0}^{1/4}$. Therefore, in the low-density limit,
the Bogoliubov description is valid, and we expect a mean-field BEC to provide a good description of the ground state.

\section{Wigner crystal state}
\label{sec4}

The first example of a strongly-correlated bosonic state we consider is the Wigner crystal (WC), proposed by Berg \textit{et al.} \cite{Berg2012}, for the Rashba SOC case. The state is constructed by dividing the volume (area) in an array of identical rectangular boxes of
size $L_x$, $L_y$, and putting each particle in a different box.
In contrast to a mean-field BEC state, the interaction energy of WC is zero, as particles completely avoid one another.
This comes at the cost of higher kinetic energy, as single-particle states are localized in boxes,
compared to a BEC, where occupied states extend throughout the entire volume. 
 
To calculate the energy per particle of the WC state, it suffices to solve for the ground state of a single particle
in a box. The calculation for the case of quartic dispersion is shown in Appendix A, and for Hamiltonian
(\ref{H_spinless}), it gives $E_g(L_x,L_y)=1.285 (\pi/L_x)^4+(\pi/L_y)^2$, where $L_x$ and $L_y$ are the length
and the width of the box. It is clear that
at low densities (small $\bm{k}$), the kinetic energy is ``cheaper'' in the $x$ than in the $y$ direction. This means that
we can lower the energy by deforming the box such that it is shorter in $x$ and longer in $y$ direction, while keeping the total volume of the box ($V=L_x L_y$), and the density ($n=1/V$) fixed.
We find the ratio $L_y/L_x$ which minimizes $E_g(L_x,L_y)$ is $L_y/L_x=0.340 \ n^{-1/3}$ and the ground-state
energy per particle is $E_g= 43.5 \ n^{4/3}$. Indeed, for the spinless case, the WC has lower energy than a mean-field BEC at low densities. By numerically solving the corresponding spinful problem [Eq.(\ref{H_spinful})], we have 
checked that the energy per particle is identical to the spinless case in the large $L_x$ limit. 

The WC state obviously has lower energy than a mean-field BEC at low densities, however it is a crystalline state 
which breaks translational symmetry. While this is expected to happen in low-density systems with long-range interactions, 
contact interactions typically do not favor formation of a crystal~\cite{Chen2012}. Therefore we expect a strongly-correlated state which is translationally invariant to have even lower energy than WC state.

We notice that here we considered only a particular type of a WC (rectangular lattice) and that different types
of WC, e.g. triangular-lattice crystal, could have lower energy. Still, we later show that
the Jastrow-type state has energy $\epsilon = 6.6\ n^{4/3}$, which is smaller than our WC state
by a factor of $7$, and we do not believe different types of WC can achieve such low energies.


%
%

\section{Strongly-correlated gas in the lower band}
\label{sec5}

\subsection{Non-interacting Fermi gas}

The Wigner crystal example motivates us to look for other strongly correlated states, constructed out of lowest band wave-functions. One natural way to build correlations is to write down wave-functions where bosons avoid one another at short distances. To see how this lowers the energy, consider a non-interacting Fermi gas in the single-particle dispersion of Eq.~(\ref{H_spinless}). The density of states corresponding to (\ref{H_spinless}) is $dn/dE = (3/2)/(2\pi)^{3/2} \times
\Gamma(5/4)/\Gamma(7/4)\ E^{-1/4}$, where $\Gamma(x)$ is the gamma function. The energy per particle in the non-interacting Fermi gas is then $\epsilon = 6.84 \ n^{4/3}$, which is indeed lower than a mean-field BEC ($\epsilon = gn/2$), and the Wigner crystal at low densities. 

It is well known that in a low-density $1$D system with contact interactions, when the contact interactions dominate the kinetic energy, the Fermi gas 
has lower energy than the mean-field BEC at the same density. This leads to ``fermionization'' 
of bosons, and the formation of a Tonks-Girardeau gas~\cite{Girardeau1960}. 

We now compute the ground state energy of several appropriately symmetrized fermionic wave-functions. We first consider the spinless case, and then generalize our results to include spin. 

\subsection{Spinless system}

\subsubsection{``Fermionized'' many-body states}
\label{sec5_2}

The ground state of a non-interacting Fermi gas in the Hamiltonian (\ref{H_spinless}) has
the following momentum distribution widths: $\Delta k_x = (4 E_{\rm F})^{1/4} \sim n^{1/3}$,
$\Delta k_y = E_{\rm F}^{1/2} \sim n^{2/3}$, where $E_{\rm F}$ is the Fermi energy. This means that, at low
densities,
the energy is minimized by broadening the distribution in the direction where kinetic energy 
is ``cheap'' ($x$ direction) and squeezing
it in the direction where energy is expensive ($y$ direction).
The WC state discussed above has the same property: $\Delta k_x \sim 1/L_x \sim n^{1/3}$ and $\Delta k_y \sim 1/L_y \sim n^{2/3}$.

To construct more general strongly-correlated bosonic wave-functions for the spinless gas, 
we take a fermionic state with the property 
$\Delta k_x \sim n^{1/3}$, 
$\Delta k_y \sim n^{2/3}$ and construct corresponding Bose wave-functions: 
for example $\psi_{B}=|\psi_F|$, $\psi_{B}=\psi_{F}^{2}$ or $\psi_{B}=|\psi_F|^2$. This way we obtain
a symmetric bosonic wave-function which  obeys $\Delta k_x \sim n^{1/3}$, 
$\Delta k_y \sim n^{2/3}$, and has kinetic energy $E_{\rm kin}/N \sim (\Delta k_{x})^{4}/4 + 
(\Delta k_{y})^{2} \sim n^{4/3}$, while the interaction energy is identically zero by construction. (In general, we can consider higher powers $\psi_{B}=|\psi_{F}|^{n}$ for $n>2$ or $\psi_{F}^{2n}$, for $n>1$, but these have higher energy, as discussed below.) 

The total energy is then simply given by 
\begin{equation}
E_{\rm kin}=\int d\bm{k} \ n_{\bm{k}} \ \varepsilon_{\bm{k}},
\label{nk_integral}
\end{equation}
where $\varepsilon_{\bm{k}}=k_{x}^{4}/4+k_{y}^{2}$ and $n_{\bm{k}}$ is the momentum distribution, normalized 
so that $\int d\bm{k}\ n_{\bm{k}}=N$. We compute the momentum distribution, and calculate the energy using Monte Carlo integration (see Appendix B for details).

But which fermionic wave-functions should we choose?
A natural choice is the ground state of a non-interacting Fermi gas ($\psi_{F,0}$).
$\psi_{F,0}$ is a real function of spatial coordinates, and we construct two
 bosonic trial wave-functions: $\psi_{B,\rm{abs}}=|\psi_{F,0}|$ and 
$\psi_{B,\rm{sq}}=\left( \psi_{F,0} \right)^2$.

%

In the case of the wave-function $\psi_{B,\rm{abs}}$, the integral (\ref{nk_integral})
diverges. The reason is that the first derivative of the wave-function is not 
continuous at points where $\psi_{B,\rm{abs}}=0$, which leads to a $\sim |\bm{k}|^{-5}$ 
decay of the momentum distribution for large $|\bm{k}|$. By contrast, $\psi_{B,\rm{sq}}$ has a continuous first derivative, and its momentum distribution vanishes for $|k_x| > 2k_{F,x}, |k_y|>2k_{F,y}$ ($k_{F,x}$ and $k_{F,y}$ are Fermi momenta of 
$\psi_{F,0}$ in $x$ and $y$ direction). The corresponding energy per particle is $\epsilon = 13.1\ n^{4/3}$, which is considerably lower than the WC energy.


A more exotic choice is the composite-fermion wave-function considered in Ref.~\cite{Sedrakyan2012} in the context of a $2$D Bose gas with Rashba SOC: 
\begin{equation}
\psi_{B,\rm{cf}}=\mathcal{N} \prod_{i<j} |z_i-z_j| \exp(-\sum_j |z_j|^2/4),
\end{equation}
 where $z_j=x_j/a_x + i y_j/a_y$,  $x_j$ and $y_j$ are particle coordinates, $a_x$ and $a_y$ are length-scales in $x$ and $y$ direction, and $\mathcal{N}$ is the normalization factor.
This state has been shown to be a quasi-condensate with algebraically decaying 
correlations~\cite{Girvin1987,Larkin_etal}, but it does not break time-reversal symmetry. 
In order to make wave-function have $\Delta k_x \sim n^{1/3}$, $\Delta k_y \sim n^{2/3}$, the lengths 
have to scale with density as $a_x \sim n^{-1/3}$, $a_y \sim n^{-2/3}$. Once again, the first derivative 
of $\psi_{B,\rm{cf}}$ is not continuous at points where $\psi_{B,\rm{cf}}=0$, and this leads to
$\sim |\bm{k}|^{-6}$ algebraic decay of the momentum distribution for large $|\bm{k}|$ 
rendering the integral (\ref{nk_integral}) divergent.

However, the square of $\psi_{B,\rm{cf}}$ wave-function:
\begin{equation}
\psi_{B,\rm{cf}}^{\rm sq}=\left( \psi_{B,\rm{cf}} \right)^2 
= \mathcal{N}^{\prime} \prod_{i<j} |z_i-z_j|^2 e^{-\sum_j |z_j|^2/2}
\end{equation}
is free from these problems. It is analytic and has an exponentially decaying momentum distribution for large $|\bm{k}|$.  Subsequently, the integral (\ref{nk_integral}) is convergent. We find that the choice of length scales which minimizes the total energy per particle is $a_x=0.55\ n^{-1/3}$, $a_y=0.29\ n^{-2/3}$, 
and the energy is $\epsilon= 9.2\ n^{4/3}$, which is the lowest energy of all the wave-functions considered so far. 
This state has zero condensate fraction, and is therefore \textit{not} a true Bose condensate. However it has algebraically decaying correlations $\rho(\bm{r},\bm{r}^{\prime})=\rho(|\bm{r}-\bm{r}^{\prime}|)=
\langle \hat{\psi}^{\dagger}(\bm{r}) \hat{\psi}(\bm{r}^{\prime}) \rangle \sim 1/|\bm{r}-\bm{r}^{\prime}|$ as $|\bm{r}-\bm{r}^{\prime}| \rightarrow \infty$, and is thus a quasi-condensate~\cite{Girvin1987,Larkin_etal}. We reproduced this result using our Monte Carlo approach (see Fig.~$1$).
\begin{figure}[h]
\centerline{
\mbox{\includegraphics[width=0.75\columnwidth]{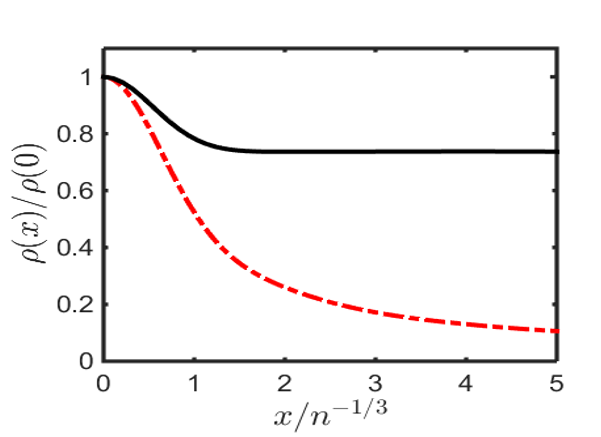}}
}
\caption{
We show the corelation function $\rho(\bm{r})$ for two different states: (a) composite-fermion state
$\psi_{B,\rm{cf}}^{\rm sq}$ (red dash-dotted line) and (b) Jastrow wave-function $\psi_{\rm J}$
(black line). Here we set $y=0$ and concentrate at the dependence on $x$.
$\psi_{B,\rm{cf}}^{\rm sq}$ wave-function has algebraically decaying correlations, i.e. it has a quasi-long range order,
while $\psi_{\rm J}$ has a true long range order (see text for details). 
}\label{ODLRO}
\end{figure} 
While it is certainly possible to consider even higher powers of $\psi_{B,\rm{cf}}$, these wave-functions
have higher energy as the increasing exponents
broaden the momentum distribution of the state.

Note that even though we casually refer to the states $\psi_{B,\rm{abs}}$, $\psi_{B,\rm{sq}}$, $\psi_{B,\rm{cf}}$, $\psi_{B,\rm{cf}}^{\rm sq}$, etc. as ``fermionized,'' the issue of fermionization is a subtle one. Strictly speaking, our ability to express a bosonic ground state wave-function in terms of properly symmetrized fermionic  wave-functions does not necessarily imply that low-energy excitations of this state have fermionic statistics. To elucidate the nature of a bosonic state in two dimensions written in terms of fermionic fields (which can always be done even for trivial ground states), one has to consider a gauge theory, e.g., either arising from a parton construction or Chern-Simons flux attachment (such as implemented by Sedrakyan et al.~\cite{Sedrakyan2012} for the bosonic Rashba model). On the other hand, there usually exists no simple way to write the corresponding many-body wave-function, which would faithfully describe gauge fluctuations, and those may have important and qualitative effects on conclusions of a na{\"\i}ve mean-field theory. For example, the many-body wave-function $\psi_{B,\rm{cf}}$ is a natural mean-field description of a ``fermionized'' state, where  fermions, obtained from original bosons via Chern-Simons flux attachment, form the integer $\nu =1$ quantum Hall state. As discussed above, the symmetrized bosonic wave-function, $\psi_{B,\rm{cf}}$, does not have a long-range order and hence appears to describe a strongly-correlated liquid state with algebraic correlations or equivalently a quasi-condensate. However, the (more general) Chern-Simons gauge-theory of the ``fermionized'' state yields a different conclusion~\cite{Jason_etal}: integrating out fermions produces another Chern-Simons term, which exactly cancels the term associated with statistical transmutation, and what is left is a gapless (Maxwell) theory. It corresponds to a Goldstone mode and indicates broken symmetry, or in other words a true condensate with long-range order. In fact, the state proposed by Sedrakyan et al.~\cite{Sedrakyan2012} belongs to this category and is a strongly-correlated BEC, rather than an exotic Bose liquid. 

All in all,  the  field-theoretical approach based on true fermionization of bosonic fields and the variational approach involving ``fermionized'' wave-functions  are not equivalent. The former provides more insight into the nature of excitations, but does not easily allow for a quantitative analysis. On the contrary, the latter can be used for explicit calculations of energy and other observables, but it does not easily elucidate the nature of low-energy excitations. One strategy here is to start with the variational approach and explore field-theoretical description, if any, of a ``fermionized'' mean-field state, if such indeed comes out as the lowest-energy trial state for a given Hamiltonian. This however does not seem to happen in our case, as discussed below.

\subsubsection{Jastrow Ansatz for a strongly-correlated BEC -- the winner}
\label{sec5_3}

One advantage of using ``fermionized'' wave-functions to approximately describe a ground state of interacting bosons is that they immediately minimize the interaction energy for any contact interaction (for spinless bosons), by the virtue of the simple fact that two fermions can not occur in the same point. However, there exist infinitely many wave-functions that accomplish the same, without relying on any fermionic analogy. Related constructions have been discussed in the literature, notably in the context of strongly-correlated BEC in Helium-4. Inspired by these previous studies, we now consider a Jastrow \textit{ansatz}~\cite{Jastrow1955} of the following form:
\begin{equation}
\psi_{\rm J} = \mathcal{N}^{\prime \prime} \prod_{i<j}^{N} \phi(\bm{r}_i-\bm{r}_j),  
\label{Jastrow}
\end{equation}
\begin{equation*}
\phi(\bm{r}) = 1 - e^{-\left(x^2/b_{x}^2+y^2/b_{y}^2\right)},  
\end{equation*}
where $\mathcal{N}^{\prime \prime}$ is the normalization, and $b_x$, $b_y$ are parameters describing correlation 
length-scale in $x$ and $y$ direction.
The density, $n=N/V$ is another important parameter of wave-function (\ref{Jastrow}). Jastrow-type
wave-functions are generally very good at capturing the behavior of Bose gases ranging from
small to large scattering lengths, i.e. from a weakly interacting to unitary regime~\cite{Rossi2014}.
A key difference here is that while usually, the Jastrow form is used to capture the short distance structure of the two-body wave-function on length scales comparable to the true atomic potential, here we work in a regime where $b_x$ and $b_y$ are on the order of the inter particle spacing, thus much larger than the scattering length. Our ansatz is therefore phenomenological in nature, and does not stem from a microscopic calculation of the two-body problem. 

As with the previously considered trial states, the Jastrow wave-function has the property that its interaction
energy is zero. By choosing $b_x \sim n^{-1/3}$, $b_y \sim n^{-2/3}$ we can  
``squeeze'' the system in the $x$ and ``stretch'' it in the $y$ direction, so that $\epsilon \sim n^{4/3}$. We find the optimal parameter values are $b_x = 0.66\ n^{-1/3}$, 
$b_y=0.29\ n^{-2/3}$, and the energy
is $\epsilon = 6.6\ n^{4/3}$. The Jastrow wave-function therefore has even lower energy than the composite-fermion wave-function $\psi_{B,\rm{cf}}^{\rm sq}$.
In Fig.~\ref{ODLRO}, we plot the single-particle density matrix $\rho(\bm{r}, \bm{r}^{\prime})$ as a function of 
$|\bm{r} - \bm{r}^{\prime}|$ corresponding to $\psi_{B,\rm{cf}}^{\rm sq}$ and the Jastrow wave-function found above. Indeed the Jastrow form has true long range order, and describes a Bose-condensate with condensate fraction $n_0/n = 0.74$. 

We therefore conclude that for the spinless Hamiltonian [Eq.(\ref{H_spinless})], although bosons can lower their energy by developing short range correlations, the correlations are not strong enough to completely destroy BEC at zero temperature. 

The \textit{ansatz} wave-functions that we considered all have the property that $\psi=0$ when $\bm{r}_i=\bm{r}_j$
which means $E_{\rm int}=0$. While this should be true in the $n \rightarrow 0$ limit,
at finite densities we expect the interaction energy not to be strictly zero. In Appendix B,
we estimate that for small densities $E_{\rm int}/N \sim n^{5/3}/g$, which means that $E_{\rm int}/E_{\rm kin} \sim n^{1/3}$.
Therefore, as in the Lieb-Liniger gas~\cite{Lieb1963}, at low densities $E_{\rm int} \ll E_{\rm kin}$. 

\subsection{Spinful system}

We now turn our attention to the spinful Hamiltonian~(\ref{H}) which corresponds to the NIST SOC scheme, and ask whether our conclusions remain valid in this case. 

We start by writing the spinful state $\lvert \psi_{B,s} \rangle$ as:
\begin{equation}
\vert \psi_{B,s} \rangle = \sum_{\bm{k}_1 ... \bm{k}_N} f_B(\bm{k}_1,...,\bm{k}_N) \lvert \bm{k}_1 ... \bm{k}_N \rangle_s,
\label{psi_s}
\end{equation}
where 
\begin{equation}
\begin{split}
f_B(\bm{k}_1,...,\bm{k}_N) = & \frac{1}{V^N} \int d\bm{r}_1 ... d\bm{r}_N \ \psi_B(\bm{r}_1,...,\bm{r}_N) \\
& \times e^{-i(\bm{k}_1 \cdot \bm{r}_1 + ... + \bm{k}_N \cdot \bm{r}_N)},
\label{momentum_rep}
\end{split}
\end{equation} 
is the Fourier transform of the spinless wave-functions $\psi_{B}$ considered above. Here $\lvert \bm{k}_1 ... \bm{k}_N \rangle_s = \lvert \bm{k}_1 \rangle_s \otimes ... \otimes \lvert \bm{k}_N \rangle_s$,
where $\lvert \bm{k} \rangle_s$ is a \textit{lower-band} eigenstate of (\ref{H}). We therefore 
\textit{construct} a spinful state exclusively from lower-band eigenstates. A similar construction was applied in Ref.~\cite{Sedrakyan2012}.


%
%

%

%

%

At low densities, the lower-band spectrum of the spinful Hamiltonian is the same 
as the spinless dispersion (\ref{H_spinless}).
Since, by construction, the spinful state [Eq.(\ref{psi_s})] has the same momentum distribution
as the corresponding spinless state, their kinetic energy is the same (see Appendix D).

However, the more complicated question is: what is the interaction energy of the spinful state? 
Since we explicily construct the spinful many-body state only from the lower-band single-particle
states, it is impossible to satisfy $\psi(....,\bm{r}_i=\bm{r}_j,...)=0,\ \forall (i,j)$, for all the different spin 
components $\psi_{\sigma_1,...,\sigma_N}(\bm{r}_1,...,\bm{r}_N)$ [$\sigma_j \in (\uparrow, \downarrow)$]. 
Therefore, unlike in the spinless wave-functions considered previously, the interaction
energy will be finite. Still, in the low-density limit, we expect
the zero overlap condition ($\psi=0$ when $\bm{r}_i=\bm{r}_j$)
to be \textit{almost} satisfied, since the system is almost completely polarized.
We thus expect the spinful state to have a very low interaction energy.

The interaction Hamiltonian (\ref{V_int}) is diagonal in real space, and to calculate the interaction energy 
it is useful to find a real-space representation of $\vert \psi_{B,s} \rangle$. Unfortunately,
the real-space representation is quite cumbersome: there are $2^N$ spin components 
(although only $N+1$ of them are independent due to the symmetric nature of $f_B$), and expressions are difficult 
to obtain:
\begin{equation}
\begin{split}
& \psi_{\sigma_1,...,\sigma_N}(\bm{r}_1,...,\bm{r}_N)= \sum_{\bm{k}_1,...,\bm{k}_N}
f_B(\bm{k}_1,...,\bm{k}_N) \\
& \quad \times s_{\sigma_1}(\bm{k}_1)...s_{\sigma_N}(\bm{k}_N) 
e^{i(\bm{k}_1 \cdot \bm{r}_1 + ... + \bm{k}_N \cdot \bm{r}_N )},
\end{split}
\end{equation}
where $s_{\sigma_j}(\bm{k})$ are given in (\ref{spin_eig}).


In Appendix E we present the method to estimate the interaction energy, and we show that 
for wave-functions $\psi_{B,\rm{sq}}$, $\psi_{B,\rm{cf}}^{\rm sq}$, and the Jastrow wave-function
the energy is $E_{\rm int}/N \sim n^{7/3}$ at low densities.
Therefore, $E_{\rm int}/E_{\rm kin} \rightarrow 0$ when $n \rightarrow 0$ and
$E_{\rm tot}/N = (E_{\rm kin} + E_{\rm int})/N  \rightarrow 6.6\ n^{4/3}$ for the Jastrow state.

It is important to assess the validity of constructing the spinful state only from
lower-band eigenstates: we have already shown that the ground state energy cannot be greater than 
$\epsilon \sim n^{4/3}$. If there was a finite fraction $u$ of particles occupying the higher band
as $n \rightarrow 0$, then the energy would be $E/N \sim u \Delta$, where $\Delta$ is the gap between
the two bands. However, this clearly contradicts the fact that $E/N \lesssim n^{4/3}$. Therefore, 
$u \rightarrow 0$ as $n \rightarrow 0$ and the $n \rightarrow 0$ ground state will only contain
states from the lower band.

\begin{table}[ht]
	\begin{center}
		\begin{tabular}{p{5.5cm} | c r}
		Wave-function              								& $\epsilon$ 											& Section \\ 
		\hline
		Mean-field BEC 														& $gn/2$  												& \ref{sec3} \\
		Wigner crystal           									& $43.5\ n^{4/3}$  								& \ref{sec4} \\
		Absolute value of Fermi-sea w.f.					& $\infty$ 												& \ref{sec5_2} \\
		Fermi-sea w.f. squared         						& $13.1\ n^{4/3}$ 								& \ref{sec5_2}	\\
		Composite-fermion w.f.										& $\infty$ 												& \ref{sec5_2}	\\
		Composite-fermion w.f. squared 						& $9.2\ n^{4/3}$ 									& \ref{sec5_2}	\\
		Jastrow state															& $6.6\ n^{4/3}$ 									& \ref{sec5_3}	\\
		\end{tabular}
	\end{center}
	\caption{Energy per particle ($\epsilon$) of different states in the low-density limit ($n$ and $g$ are
	dimensionless density and interaction strength, respectively). Of all the wave-functions (w.f.) we consider,
	the Jastrow state	has the lowest energy. Two wave-functions (absolute value of Fermi-sea w.f. 
	and composite-fermion w.f.) have diverging expectation value of $k_{x}^{4}$ (see text for details). 
	}
\label{tab1}
\end{table}

\section{Discussion and experimental relevance}
In this paper, we considered a system of interacting bosons with a quartic single-particle dispersion. It was shown that the  low-density limit of the model hosts a strongly-correlated ground state, where the mean-field Bogoliubov state can be easily ruled out as being parametrically higher in energy than the strongly-correlated states, where bosons develop local correlations and avoid each other. 

Among the many trial states we considered, a long-range-ordered condensate described by the Jastrow wave-function 
[Eq.(\ref{Jastrow})] was found to have the lowest energy per particle of $\epsilon = 6.6\ n^{4/3}$ 
(compared to $\epsilon = gn/2$ for a mean-field BEC). This is in agreement with Ref.~\cite{QiZhou2014} where
it was argued that the ground state of system (\ref{H_spinless}) has long-range order. The condensate fraction was found to be $N_0/N=0.74$. \textit{i.e}, it is a strongly-correlated BEC with significant depletion of the condensate due to the interplay between interactions and the unusual band structure.

Importantly, the mean-field BEC and the Jastrow BEC break the same symmetry, therefore, we expect that the system continuously evolves from a weakly to a strongly-correlated BEC state from high to low-densities, without any phase transitions in between (a similar weak-to-strong-coupling cross-over can be tuned by evolving the single-particle dispersion from the usual quadratic to quartic). 

Note however that in the absence of a systematic procedure to explore many-body ground states of strongly-correlated systems, our variational-approach results are strongly suggestive, but not conclusive.
Eventually, it is experiment that would fully elucidate the nature of the ground state, and to realize our model is at the experimentalists' fingertips.

The strongly-correlated condensate we predict can be detected experimentally using a number of probes. 
For example, the suppressed condensate fraction is measurable in time-of-flight~\cite{BlochRMP}.
Another signature of strong correlations is the ratio of interaction and kinetic energy which is very small at low densities. This could be accessed via quantum quench experiments, i.e. 
the interaction parameter $g$ could be suddenly changed and the effect on the total energy of the gas 
could be measured. 
Strong local correlations can also be measured \textit{in situ} by observing the anti-bunching of bosonic atoms~\cite{Greiner2015}. 
Finally, several groups \cite{Esslinger2000,Schmiedmayer2013} have directly measured $\rho(\bm{r},\bm{r}^{\prime})$.
This would give information about the condensate fraction, and the type of order present in the gas.

We can estimate the density below which strong correlations become energetically favourable by equating
mean-field-BEC and Jastrow-state energies (see Table \ref{tab1}). In the case of $^{87}$Rb with a $z$-direction confinement frequency $\omega_z=2\pi \times 4000\ {\rm Hz}$, this gives $n \approx 10^{-6}\ k_{L}^{2} \approx 
6 \times 10^{7}\ {\rm m}^{-2}$ ($k_L = 2\pi/\lambda$, where $\lambda \approx 800\ {\rm nm}$~\cite{Spielman2009}), which is much lower than typical densities in cold-atom experiments studying $2$D systems ($n \sim 10^{13}\ {\rm m}^{-2}$~\cite{Dalibard2006}). However, using Feshbach
resonances to increase $g$, it is possible to make strong correlations favourable at considerably 
higher densities, up to $n \approx 0.004\ k_{L}^2 \approx 2 \times 10^{11}\ {\rm m}^{-2}$, which could be achieved experimentally. At densities higher than this the dispersion in $x$-direction cannot be approximated by a quartic term anymore, and higher-order terms have to be included.

\section{Acknowledgements} This work was supported by Army Research Office (ARO) MURI (J.R. and S.N.), 
the Physics Frontier Center at the Joint Quantum Institute funded by the National Science Fundation (NSF) (S.N.), 
Air Force Office of Scientific Research (AFOSR) MURI (S.N.), and US-ARO, Australian Research Council, and Simons Foundation (V.G.). 
SN would like to thank the Department of Energy's Institute for Nuclear Theory at the University of Washington for its hospitality and the Department of Energy for partial support during the completion of this work. We are grateful to Jason Alicea, Tin-Lun Ho, Alex Kamenev, Olexei Motrunich, and Tigran Sedrakyan for useful discussions.

\begin{appendix}

\section{Box-potential ground state in a system with quartic dispersion}

Here we show how to find a spectrum of a particle in box potential
with Hamiltonian $H_4=k_{x}^{4}=\partial_{x}^{4}$. While $H_4$ is similar
to the usual quadratic dispersion in a sense that both are diagonal in momentum space,
there is one fundamental difference: in a system with $H_4$, not only the wave-function, but
also its first derivative 
has to be continuous for the wave-function to have finite energy expectation value. 

If we choose a box of length $L$ and $-L/2 < x < L/2$, then the
boundary conditions are $\psi(-L/2)=\psi(L/2)=0$ and $\partial_x \psi(-L/2)=\partial_x \psi(L/2)=0$.
Solutions of the equation $\partial_{x}^{4} \psi = E \psi$ are $\exp(kx)$, $\exp(-kx)$,
$\exp(ikx)$ and $\exp(-ikx)$, where $k=E^{1/4}$. 
Since $H_4$ is symmetric under inversion ($x \rightarrow -x$),
we expect a symmetric ground state:
\begin{equation}
\psi_0(x)=a_1 \left( e^{kx} + e^{-kx} \right) + a_2 \cos(kx),
\end{equation}
where $a_1$, $a_2$ are coefficients that have to be determined.
Boundary conditions then require $\tan(kL/2) = -\tanh(kL/2)$, which can be solved 
graphically: in the ground state $kL=4.730$ and $E_0=5.140\ (\pi/L)^4$.
The ratio of coefficients is $a_2/a_1=15.06$.

\section{Monte-Carlo calculations}

Here we describe Monte-Carlo methods we used to calculate the kinetic energy of various trial 
wave-functions. We were primarily interested in finding the expectation value of Hamiltonian (\ref{H_spinless})
and for analytic wave-functions this can be done in two ways. As in standard Variational Monte Carlo techniques, the first step is to sample the ``local energy'', 
$E_{\rm loc} = (\hat{H} \psi)/\psi$ \cite{KentThesis}:
\begin{equation}
\begin{split}
E_{\rm kin} & = \frac{\int \psi^{*} \hat{H} \psi d\bm{R} }{\int |\psi|^2 d\bm{R}} \\
& = \frac{\int |\psi|^2 \frac{\hat{H} \psi}{\psi} d\bm{R} }{\int |\psi|^2 d\bm{R}},
\end{split}
\end{equation}
where $\hat{H} = \sum_j \partial_{x_j}^{4}/4 - \partial_{y_j}^{2}$, and $d\bm{R}=d\bm{r}_1 d\bm{r}_2 \cdots d\bm{r}_N$. 
We then use a Metropolis algorithm to sample the local energy with probability distribution 
$P(\bm{R})=|\psi(\bm{R})|^2 / \int |\psi(\bm{R}^{\prime})|^2 d\bm{R}^{\prime}$. 

In the case of non-analytic wave-functions like $\psi_{B,\rm{abs}}$ and $\psi_{B,\rm{cf}}$ the expectation
value of $\partial_{x_j}^{4}$ cannot be calculated this way because a finite energy is associated with
points which have discontinuous derivatives of $\psi$. 
The correct method in that case is to first calculate the momentum distribution
of the state, and then compute the expectation value of $\varepsilon_k = k_{x}^4/4 + k_{y}^2$. We calculate 
the momentum distribution $n(\bm{k})$ using:
\begin{equation}
n(\bm{k}) = N \int d\bm{r}_2 \cdots d\bm{r}_N |f(\bm{k},\bm{r}_2,...,\bm{r}_N)|^2,
\end{equation}
\begin{equation*}
f(\bm{k},\bm{r}_2,...,\bm{r}_N) = \frac{1}{2\pi} \int d\bm{r}_1 e^{-i\bm{k} \cdot \bm{r}_1}
\psi(\bm{r}_1,...,\bm{r}_N),   
\end{equation*}
where we chose the following normalization: $\int |\psi(\bm{R})|^2 d\bm{R} = 1$
and $\int n(\bm{k}) d\bm{k} = N$.
This can be written in the form suitable for Metropolis importance sampling:
\begin{equation}
n(\bm{k})= N \int d\bm{R} |\psi(\bm{R})|^2 |f_{\mathcal{N}}(\bm{k},\bm{r}_2,...,\bm{r}_N)|^2,
\end{equation}
\begin{equation*}
\begin{split}
f_{\mathcal{N}}(\bm{k},\bm{r}_2,...,\bm{r}_N) & = \frac{1}{2\pi} \int d\bm{r}_1 e^{-i\bm{k} \cdot \bm{r}_1} 
\psi_{\mathcal{N}}(\bm{r}_1,...,\bm{r}_N), \\
\psi_{\mathcal{N}}(\bm{r}_1,...,\bm{r}_N) & = \frac{\psi(\bm{r}_1,...,\bm{r}_N) }
{\sqrt{\int d\bm{r}^{\prime} |\psi(\bm{r}^{\prime},\bm{r}_2,...,\bm{r}_N)|^2}}.
\end{split}
\end{equation*}
For analytic wave-functions, both methods produce the same result.   

The wave-functions we considered all have the property of being very anisotropic, i.e. they are given in terms of
length-scales $a_x$, $a_y$ where $a_x \ll a_y$. The best way to do calclations is then to
rescale the coordinates so that in the new units $a_x \sim a_y \sim 1$. For example, in the case
of $\psi_{B,{\rm cf}}^{\rm sq}$ we 
first calculate expectation values  $\alpha_4=\langle k_{x}^{4} \rangle$ and $\alpha_2=\langle k_{y}^{2} \rangle$ for the wave-function with $a_x = a_y = 1$ (rescaled wave-function). The
expectation values corresponding to the wave-function in the original units are then simply
$\alpha_4/a_{x}^{4}$ and $\alpha_2/a_{y}^{2}$, respectively. In the end, we minimize 
$E(a_x,a_y)=\langle k_{x}^{4} \rangle/4 + \langle k_{y}^{2} \rangle = \alpha_4/(4a_{x}^{4}) + \alpha_2/a_{y}^{2}$,
while keeping the density $n = 1/(2\pi a_x a_y)$ constant. 

In the case of wave-functions where we can apply periodic boundary condition ($\psi_{B,{\rm abs}}$,
$\psi_{B,{\rm sq}}$, and $\psi_{\rm J}$) we did calculations with $N=400$ particles. 
However, in the case of composite-fermion wave-functions ($\psi_{B,{\rm cf}}$ and $\psi_{B,{\rm cf}}^{\rm sq}$) 
we did calculations with $N=1600$ particles.
There the density of a wave-function with finite number of particles has a form of a droplet with 
radius $R = \sqrt{2(N-1)}$ (when $a_x=a_y=1$) and the presence of the boundary increases the value
of finite-size correction. Larger system sizes were therefore necessary.


\section{Estimating the interaction energy of a spinless gas at small, but finite densities}

In order to estimate the interaction energy at small densities,
we can make a simple order-of-magnitude calculation: we choose some coordinates 
$\bm{\lambda}=(\bm{r}_3,...,\bm{r}_N)$
and we keep them fixed (see Appendix E for more details). Now we can write the wave-function as $\psi(\bm{r}_1,\bm{r}_2;\bm{\lambda})$,
and we can define parameter $C$ as a measure of a $\bm{r}_1=\bm{r}_2$ wave-function amplitude: 
\begin{equation}
C=V \frac{ \int d\bm{r} |\psi(\bm{r},\bm{r})| }{\int d\bm{r}_1 d\bm{r}_2 
|\psi(\bm{r}_1,\bm{r}_2)| },
\end{equation}
where $V$ is the volume.
We first estimate the kinetic energy: $E_{\rm kin}(C) \sim n^{4/3} (1-C)$. The reasoning is that for $C=0$, $E_{\rm kin} \sim n^{4/3}$. When $C=1$, the gas is not correlated and $E_{\rm kin} \approx 0$. Moreover, $E_{\rm kin}$ should not have an extremum around $C=0$,
and therefore should be linear in $C$ in that region.

The interaction energy is $E_{\rm int} \sim N^2 g \int d\bm{r} |\psi(\bm{r},\bm{r})|^2 \sim g n C^2$.
We minimize $E_{\rm kin}+E_{\rm int}$ with respect to $C$ and the optimal $C$ is $C \sim n^{1/3}/g$,
and $E_{\rm int} \sim n^{5/3}/g$. This means $E_{\rm int}/E_{\rm kin} \sim n^{1/3}$, that is the kinetic
energy is a dominant part at low denities. The same reasoning gives the 
correct density scaling of $E_{\rm int}$ in the low-density regime of a $1$D Lieb-Liniger gas.

\section{Kinetic energy of a many-body wave-function}

Here we show that spinful wave-function constructed in Eq.(\ref{psi_s}) has the same kinetic
energy as the corresponding spinless wave-function.

We consider the following state:
\begin{equation}
\lvert \psi \rangle = \sum_{\bm{k}_1 ... \bm{k}_N} f_B(\bm{k}_1,...,\bm{k}_N) \lvert \bm{k}_1 ... \bm{k}_N \rangle,
\label{psi_appD}
\end{equation}
where $f_B$ is normalized: $\sum_{\bm{k}_1,...,\bm{k}_N} |f_B|^2 = 1$, and $\lvert \bm{k}_1 ... \bm{k}_N \rangle=\lvert \bm{k}_1 \rangle \otimes ... \otimes \lvert \bm{k}_N \rangle$ 
is an orthonormal momentum-eigenstate basis. Here state $\lvert \bm{k} \rangle$ can describe either a spinless or
spinful [Eq.(\ref{spin_eig})] single-particle momentum eigenstate.  
The state $\lvert \bm{k}_1 ... \bm{k}_N \rangle$ is an eigenstate of kinetic energy operator
with energy $\varepsilon_{\bm{k}_1 ... \bm{k}_N}=\varepsilon_{\bm{k}_1}+...+\varepsilon_{\bm{k}_N}$.
Therefore
\begin{equation}
\begin{split}
E_{\rm kin} & =\sum_{\bm{k}_1 ... \bm{k}_N} \varepsilon_{\bm{k}_1 ... \bm{k}_N} |f_B(\bm{k}_1,...,\bm{k}_N)|^2 \\
& = \sum_{\bm{k}_1 ... \bm{k}_N} \left( \varepsilon_{\bm{k}_1}+...+\varepsilon_{\bm{k}_N} \right) |f_B(\bm{k}_1,...,\bm{k}_N)|^2 \\
& = \sum_{\bm{k}_1} \varepsilon_{\bm{k}_1} \sum_{\bm{k}_2 ... \bm{k}_N} |f_B(\bm{k}_1,...,\bm{k}_N)|^2 + \cdots \\
& = \sum_{\bm{k}} \varepsilon_{\bm{k}} n_{\bm{k}},
\end{split}
\end{equation}
where we assumed $f_B$ is symmetric with respect to particle exchange and 
$n_{\bm{k}}$ is the single-particle momentum distribution:
\begin{equation}
n_{\bm{k}}= N \sum_{\bm{k}_2 ... \bm{k}_N} |f_B(\bm{k},\bm{k}_2,...,\bm{k}_N)|^2.
\end{equation}
It is clear that kinetic energy does not depend on whether $\lvert \psi \rangle$ [Eq.(\ref{psi_appD})]
describes a spinless or spinful state, as long their momentum representation $f_B$ and dispersion
$\varepsilon_{\bm{k}}$ are the same.

\section{Estimating spinful state interaction energy}

The spinful state is defined as
\begin{equation}
\lvert \psi_{B,s} \rangle = \sum_{\bm{k}_1 ... \bm{k}_N} f_B(\bm{k}_1,...,\bm{k}_N) \lvert \bm{k}_1 ... \bm{k}_N \rangle_s,
\end{equation}
where $\lvert \bm{k}_1 ... \bm{k}_N \rangle_s=\lvert \bm{k}_1 \rangle_s \otimes ... \otimes \lvert \bm{k}_N \rangle_s$ 
and $\lvert \bm{k} \rangle_s$ is a lower-band single-particle state.

The real-space representation of $\lvert \bm{k}_1 ... \bm{k}_N \rangle_s$ is
\begin{equation*}
\begin{bmatrix}
\langle \bm{r}_1 ... \bm{r}_N; \uparrow \uparrow ... \uparrow \rvert \bm{k}_1 ... \bm{k}_N \rangle_s \\ 
\langle \bm{r}_1 ... \bm{r}_N; \uparrow \uparrow ... \downarrow \rvert \bm{k}_1 ... \bm{k}_N \rangle_s \\ 
. \\
. \\
. \\
\langle \bm{r}_1 ... \bm{r}_N; \downarrow \downarrow ... \uparrow \rvert \bm{k}_1 ... \bm{k}_N \rangle_s \\ 
\langle \bm{r}_1 ... \bm{r}_N; \downarrow \downarrow ... \downarrow \rvert \bm{k}_1 ... \bm{k}_N \rangle_s 
\end{bmatrix}
=
\begin{bmatrix}
s_{\uparrow}(\bm{k}_1) s_{\uparrow}(\bm{k}_2) ... s_{\uparrow}(\bm{k}_N) \\
s_{\uparrow}(\bm{k}_1) s_{\uparrow}(\bm{k}_2) ... s_{\downarrow}(\bm{k}_N) \\
. \\
. \\
. \\
s_{\downarrow}(\bm{k}_1) s_{\downarrow}(\bm{k}_2) ... s_{\uparrow}(\bm{k}_N) \\
s_{\downarrow}(\bm{k}_1) s_{\downarrow}(\bm{k}_2) ... s_{\downarrow}(\bm{k}_N)
\end{bmatrix}
\end{equation*}
\begin{equation}
\times e^{i(\bm{k}_1 \cdot \bm{r}_1 + ... + \bm{k}_N \cdot \bm{r}_N)},
\end{equation}
where $s_{\uparrow}(\bm{k})$, $s_{\downarrow}(\bm{k})$ are given in eq.(\ref{spin_eig}).
The real-space representation of the spinful wave-function is then
\begin{equation*}
\begin{bmatrix}
\psi_{\uparrow \uparrow \cdots \uparrow}(\bm{r}_1, \cdots, \bm{r}_N) \\
\psi_{\uparrow \uparrow \cdots \downarrow}(\bm{r}_1, \cdots, \bm{r}_N) \\
. \\
. \\
. \\
\psi_{\downarrow \downarrow \cdots \uparrow}(\bm{r}_1, \cdots, \bm{r}_N) \\
\psi_{\downarrow \downarrow \cdots \downarrow}(\bm{r}_1, \cdots, \bm{r}_N)
\end{bmatrix}
=  \sum_{\bm{k}_1 ... \bm{k}_N} f_B(\bm{k}_1,...,\bm{k}_N)
\end{equation*}
\begin{equation}
\times
\begin{bmatrix}
s_{\uparrow}(\bm{k}_1) s_{\uparrow}(\bm{k}_2) ... s_{\uparrow}(\bm{k}_N) \\
s_{\uparrow}(\bm{k}_1) s_{\uparrow}(\bm{k}_2) ... s_{\downarrow}(\bm{k}_N) \\
. \\
. \\
. \\
s_{\downarrow}(\bm{k}_1) s_{\downarrow}(\bm{k}_2) ... s_{\uparrow}(\bm{k}_N) \\
s_{\downarrow}(\bm{k}_1) s_{\downarrow}(\bm{k}_2) ... s_{\downarrow}(\bm{k}_N)
\end{bmatrix}
e^{i(\bm{k}_1 \cdot \bm{r}_1 + ... + \bm{k}_N \cdot \bm{r}_N)}.
\label{psi_r}
\end{equation}
We are interested in the low-density regime and there $|\bm{k}| \ll 1$. We can
then expand spin coefficients as:
\begin{equation}
\begin{split}
s_{\uparrow}(\bm{k}) & =-\frac{1}{\sqrt{2}} \left( 1 -\frac{k_x}{2}-\frac{k_{x}^2}{8} + \mathcal{O}(k_{x}^{3})\right), \\
s_{\downarrow}(\bm{k}) & =\frac{1}{\sqrt{2}} \left( 1 +\frac{k_x}{2}-\frac{k_{x}^2}{8} + \mathcal{O}(k_{x}^{3}) \right)
\end{split}
\end{equation}
Also, we can replace $k_x$ with $-i \partial_x$. For example,
the $\psi_{\downarrow \cdots \downarrow}$ component is then:
\begin{equation}
\begin{split}
\psi_{\downarrow \cdots \downarrow} & = \left( \frac{1}{\sqrt{2}}  \right)^N 
\left( 1 - \frac{i\partial_{x_1}}{2}  + \frac{ \partial_{x_1}^{2}}{8} + ...\right) \\
&\times \cdots \times \left( 1 - \frac{i \partial_{x_N}}{2} + \frac{ \partial_{x_N}^{2}}{8} + ...\right) 
\psi_B(\bm{r}_1, \cdots, \bm{r}_N),
\end{split}
\label{psi_down}
\end{equation}
since by definition of $f_B$ (Eq.(\ref{momentum_rep})):
\begin{equation}
\begin{split}
\psi_B(\bm{r}_1, \cdots, \bm{r}_N) = & \sum_{\bm{k}_1 \cdots \bm{k}_N} f_B(\bm{k}_1, \cdots,\bm{k}_N) \\
& \times e^{i\left( \bm{k}_1 \cdot \bm{r}_1 + \cdots + \bm{k}_N \cdot \bm{r}_N \right)}.
\end{split}
\end{equation}

The strategy for calculating the interacting energy is: 
(a) we concentrate on expectation value of $V_{12}=g \delta(\bm{r}_1-\bm{r}_2)$ (the total interaction 
energy will then be $N(N-1)/2$ times that value). 
(b) We first calculate the contribution to interaction energy coming from $\psi_{\downarrow \cdots \downarrow}$ 
component.
(c) We show that all other spin components give approximately the same contribution.

{\it Estimating $\psi_{\downarrow \cdots \downarrow}$ contribution.---}
The idea is to choose some random values for coordinates $\bm{r}_3, \cdots, \bm{r}_N$ 
and keep them fixed [we define $\bm{\lambda}=(\bm{r}_3, \cdots, \bm{r}_N)$]. 
This way we get a two-body wave-function from which it is easy to calculate $\langle V_{12} \rangle$.
Later we show that almost any choice of $\bm{\lambda}$ gives the same value of $\langle V_{12} \rangle$.

We start by defining
\begin{equation}
\begin{split}
\Phi(\bm{r}_1,\bm{r}_2; \bm{\lambda}) & = \mathcal{N}(\bm{\lambda}) 
\left( 1 - \frac{i\partial_{x_1}}{2}  
+ \frac{ \partial_{x_1}^{2}}{8} + ...\right) \\
& \times \left( 1 - \frac{i \partial_{x_2}}{2} + \frac{ \partial_{x_2}^{2}}{8} + ...\right) 
\chi(\bm{r}_1,\bm{r}_2; \bm{\lambda}),
\end{split}
\end{equation}
where
\begin{equation}
\begin{split}
\chi(\bm{r}_1,\bm{r}_2; \bm{\lambda})= & \left( 1 - \frac{i\partial_{x_3}}{2}  
+ \frac{ \partial_{x_3}^{2}}{8} + \cdots \right) \\
& \times \cdots \times \left( 1 - \frac{i \partial_{x_N}}{2} + \frac{ \partial_{x_N}^{2}}{8} + \cdots \right) \psi_B,
\end{split}
\end{equation}
and $\mathcal{N}(\bm{\lambda})$ is such that
\begin{equation}
\int d\bm{r}_1 d\bm{r}_2 \  |\Phi(\bm{r}_1,\bm{r}_2; \bm{\lambda})|^2 = 1.
\end{equation}
We notice that $\Phi(\bm{r}_1,\bm{r}_2; \bm{\lambda})$ is simply $\psi_{\downarrow \cdots \downarrow}$ with
a different normalization [see Eq.(\ref{psi_down})].

If $\psi_B$ is analytical, we can Taylor-expand
around $\bm{r}_1-\bm{r}_2=0$:
\begin{equation}
\begin{split}
\psi_B = & \frac{1}{2} a_{xx} \left( x_1-x_2 \right)^2 + \frac{1}{2} a_{yy} \left( y_1-y_2 \right)^2 \\
& + a_{xy} \left( x_1-x_2 \right) \left( y_1-y_2 \right) + \cdots,
\end{split}
\end{equation}
where we assume $\psi_B=0$ when $\bm{r}_1-\bm{r}_2=0$ and $a_{ij}=a_{ij}(\bm{r}_{\rm cm},\bm{\lambda})$,
where $\bm{r}_{\rm cm}=\bm{r}_1+\bm{r}_2$.
$\chi$ retains the same structure, but with different coefficients:
\begin{equation}
\begin{split}
\chi = & \frac{1}{2} b_{xx} \left( x_1-x_2 \right)^2 + \frac{1}{2} b_{yy} \left( y_1-y_2 \right)^2 \\
& + b_{xy} \left( x_1-x_2 \right) \left( y_1-y_2 \right) + \cdots,
\end{split}
\end{equation}
However, once we act on $\chi$ with $s_{\downarrow}(-i\partial_{x_1}) s_{\downarrow}(-i\partial_{x_2})$,
function $\Phi$ will have non-zero value for $\bm{r}_1=\bm{r}_2$ which will give rise to finite interaction 
energy.

Let us make the change of variables: $x_r=x_1-x_2$, $x_{\rm cm}=x_1+x_2$.
Then:
\begin{equation*}
\begin{split}
\Phi|_{\bm{r}_1 = \bm{r}_2} \approx & \left( 1 - \frac{i\partial_{x_1}}{2}  
+ \frac{ \partial_{x_1}^{2}}{8} \right) \\
& \times \left( 1 - \frac{i \partial_{x_2}}{2} + \frac{ \partial_{x_2}^{2}}{8} \right) \chi(\bm{r}_1,\bm{r}_2; \bm{\lambda})|_{\bm{r}_1 = \bm{r}_2}
\end{split}
\end{equation*}
\begin{equation*}
\begin{split}
 = & \left[ 1 - \frac{i}{2} (\partial_{x_r}+\partial_{x_{\rm cm}}) 
+ \frac{1}{8}(\partial_{x_r}+\partial_{x_{\rm cm}})^2 \right] \\
& \times \left[ 1 - \frac{i}{2} (-\partial_{x_r}+\partial_{x_{\rm cm}}) 
+ \frac{1}{8}(-\partial_{x_r}+\partial_{x_{\rm cm}})^2 \right] \\
& \times \chi(\bm{r}_1,\bm{r}_2; \bm{\lambda})|_{\bm{r}_1 = \bm{r}_2} \\
= & \left( 1 + \frac{1}{2} \partial_{x_r}^{2} + \cdots \right) 
\chi(\bm{r}_1,\bm{r}_2; \bm{\lambda})|_{\bm{r}_1 = \bm{r}_2} \\
= & \frac{b_{xx}}{2},
\end{split}
\end{equation*}
We can estimate $b_{xx} \sim (\Delta k_x)^2 \bar{|\Phi|}$, where $\Delta k_x$ is the momentum width
in $x$ direction and $\bar{|\Phi|}$ is the average magnitude of $\Phi$:
\begin{equation}
\bar{|\Phi|}^2= \frac{1}{V^2} \int d\bm{r}_1 d\bm{r}_2 \ |\Phi|^2 =\frac{1}{V^2},
\end{equation}
where $V$ is the volume.
The interaction energy corresponding to $\Phi$ is then
\begin{equation}
\begin{split}
E_{12}(\Phi) &= \int d\bm{r}_1 d\bm{r}_2 \ g \delta(\bm{r}_1-\bm{r}_2) 
|\Phi(\bm{r}_1,\bm{r}_2)|^2 \\
& \sim V g (\Delta k_x)^4 \bar{|\Phi|}^2 \sim \frac{g}{V} (\Delta k_x)^4
\end{split}
\end{equation}
It is clear that a different choice of $\bm{\lambda}$ would give the same estimate. The same is true
for different spin components $\psi_{\sigma_1,...,\sigma_N}$. Therefore, when we average over
all $\lambda$ and $\{\sigma_1,...,\sigma_N\}$:
\begin{equation}
\begin{split}
\bar{E}_{12}&= \sum_{\sigma_1,...,\sigma_N} \int d\bm{\lambda}\ p_{\sigma_1,...,\sigma_N}(\bm{\lambda}) E_{12}[\Phi_{\sigma_1,...\sigma_N}(\bm{\lambda})], \\
 &p_{\sigma_1,...,\sigma_N}(\bm{\lambda}) = \int d\bm{r}_1 d\bm{r}_2 |\psi_{\sigma_1,...,\sigma_N}(\bm{r}_1,\bm{r}_2,\bm{\lambda})|^2,
\end{split}
\end{equation}
%
%
the energy is again $\bar{E}_{12} \sim \frac{g}{V} (\Delta k_x)^4$.
The total energy is then $E_{\rm int} \sim N(N-1)/2 \times \bar{E}_{12} \sim N g n (\Delta k_x)^4$ (since
there are $N(N-1)/2$ interacting pairs), 
that is $E_{\rm int}/N \sim g n (\Delta k_x)^4$. 
The states that we considered ($\psi_{B,\rm{sq}}$, $\psi_{B,\rm{cf}}^{\rm sq}$, and $\psi_{\rm J}$) have 
$\Delta k_x \sim n^{1/3}$ which leads to $E_{\rm int}/N \sim g n^{7/3}$.

The method described here works only if derivatives of $\psi_B$ are defined at points
where $\bm{r}_i=\bm{r}_j$. This is, for example, not the case with $\psi_{B,\rm{abs}}$ or $\psi_{B,\rm{cf}}$.
However, in those cases it is still possible to estimate the interaction energy: we again concentrate
on a two-body wave-function, that is, we fix $\bm{\lambda}=(\bm{r}_3, \cdots, \bm{r}_N)$ and 
we look what happens with $\psi_{\downarrow \cdots \downarrow}(\bm{r}_1,\bm{r}_2;\bm{\lambda})$. Then we can estimate
the value of $\psi(\bm{r}_1=\bm{r}_2)$ by looking into its Fourier transform.

The general conclusion is that $E_{\rm int}/N \sim g n (\Delta k_x)^{2m}$, where $m=1$ if
the first derivative of $\psi$ with respect to relative distance $\bm{r}_{ij}=\bm{r}_i-\bm{r}_j$
does not approach zero as $\bm{r}_{ij} \rightarrow 0$, $m=2$ if the first derivative approaches zero,
but the second derivative does not as $\bm{r}_{ij} \rightarrow 0$, etc.
For example, wave-functions $\psi_{B,\rm{abs}}$ or $\psi_{B,\rm{cf}}$ therefore have 
$E_{\rm int}/N \sim g n (\Delta k_x)^2$.

\end{appendix}

\bibliography{Fermionized_references}

\end{document}